\documentclass[reprint,
superscriptaddress,
 amsmath,amssymb,
 aps,
prb,
]{revtex4-2}

\usepackage{graphicx}
\usepackage{dcolumn}
\usepackage{bm}
\usepackage{color}
\usepackage{xcolor}
\usepackage{hyperref}

\begin{document}

\title{Expanding the search space of high entropy oxides and predicting synthesizability using machine learning interatomic potentials}

\author{Oliver A. Dicks}
\email[Corresponding author. Email address: ]{oliver.dicks@ubc.ca}
\affiliation{Stewart Blusson Quantum Matter Institute, University of British Columbia, Vancouver, British Columbia V6T 1Z4, Canada}

\author{Solveig S. Aamlid}
\affiliation{Stewart Blusson Quantum Matter Institute, University of British Columbia, Vancouver, British Columbia V6T 1Z4, Canada}

\author{Alannah M. Hallas}
\affiliation{Stewart Blusson Quantum Matter Institute, University of British Columbia, Vancouver, British Columbia V6T 1Z4, Canada}
\affiliation{Department of Physics \& Astronomy, University of British Columbia, Vancouver, British Columbia V6T 1Z1, Canada}
\affiliation{Canadian Institute for Advanced Research (CIFAR), Toronto, ON, M5G 1M1, Canada}
    
\author{Joerg Rottler}
\affiliation{Stewart Blusson Quantum Matter Institute, University of British Columbia, Vancouver, British Columbia V6T 1Z4, Canada}
\affiliation{Department of Physics \& Astronomy, University of British Columbia, Vancouver, British Columbia V6T 1Z1, Canada}

\date{\today}

\begin{abstract}

We propose an efficient computational methodology for predicting the synthesizability of high entropy oxides (HEOs) in a large space of possible candidate compounds.
HEOs are a growing field with an enormous potential chemical composition space, and yet the discovery of new HEOs is slow and driven by experimental trial-and-error.
In this work, we attempt to speed up this process by using a machine learned interatomic potential offering DFT-level accuracy.
Our methodology starts by identifying a set of crystal structures and elements for screening, building a large random unit cell of each composition and structure, then relaxing this structure. The most promising candidates are distinguished
based on the variance of the individual cation energies, which we introduce as our entropy descriptor, and the enthalpy of mixing, which is used as the enthalpy descriptor.
The approach is applied to tetravalent HEOs, and its validity is confirmed by comparison to alternative descriptors and DFT calculations for a set of 7 elements. The search is then extended to a set of 14 elements and three crystal structures, where it successfully identifies the only known stable 4-component HEO in the $\alpha$-PbO$_2$ structure, as well as predicting several new 5-component candidate systems. This approach can straightforwardly be applied to new sets of elements and structures, allowing for the accelerated discovery of new HEOs.


\end{abstract}

\maketitle


\section{\label{sec:Intro} Introduction}

High entropy oxides (HEOs) have garnered significant interest due to their ability to stabilize multicomponent chemistries through configurational entropy~\cite{Rost2015,zhang2019review,oses2020high,musico2020emergent,mccormack2021thermodynamics,aamlid_understanding_2023}. This class of materials displays emergent and attractive functionalities for a host of applications, such as enhanced cycling stability in reversible energy storage \cite{Sarkar2018}, room-temperature lithium superionic conductivity \cite{Brardan2016}, catalysis \cite{Sun2021}, and low thermal conductivity for thermal barrier coatings \cite{Li2019}. The possible compositional space of HEOs is enormous; if non-metals and elements with no stable isotopes are excluded, there are 65 elements left in the periodic table which could play the role of a cation in an oxide. Choosing 5 elements out of these 65 can be done more than 8 million different ways. For comparison, the inorganic crystal structure database contains just short of 320,000 entries (ICSD release 2025.1~\cite{ICSD}). This chemical composition space is even larger if off-equimolar compositions or combinations of four or six or more elements are considered. There are a limited number of reported HEOs, and the parameter space is impossibly large for experimental screening, which leaves the field ripe for computational high-throughput approaches.

Despite their promise for applications and the vastness of the compositional space, accurate computational design of synthesizable HEOs remains challenging. The primary difficulty arises from the sheer scale of possible atomic configurations which have to be evaluated. While density functional theory (DFT) has been invaluable for performing structure searches of HEOs, its computational cost limits the number of compositions and structures which can be screened, as well as the supercell size, which must be sufficiently large to model these highly disordered systems. Descriptor-based approaches, where a proxy quantity is calculated and presumed to capture the most important features, are a common route to limit computational cost~\cite{Pitike2020, dey_mixed_2024, Divilov2024}. There may be additional kinetic factors or unknown competing phases at play, further limiting the number of synthesizable HEOs, such that heuristics and perseverance remain the path to successful discoveries.

Recent advances in machine learning interatomic potentials (MLIPs) provide an avenue to overcome the cost limitations of DFT. Calculations of complex systems can be performed for a fraction of the computational cost without the need for simplifications, while retaining close-to-DFT accuracy. Universal potentials such as CHGNet~\cite{Deng2023} have already been applied to performing computational searches for novel high entropy materials~\cite{sivak_discovering_2025}. The MACE foundation model~\cite{batatia2023foundation}, which is used in this work, has demonstrated excellent performance across a range of chemistries without the need for task-specific retraining. 

Leveraging the MACE foundation model, we propose a set of methods to perform high-throughput structure searches of HEOs combined with accurate predictions of their synthesizability. Tetravalent HEOs are used as a model system, since we have access to previous computational and experimental data for comparison. The proposed approach can be applied to other crystal structures and sets of elements to allow us to identify trends and heuristics that in combination with experimental efforts can target synthesizable compositions within the vast chemical space of high entropy oxides.

\section{\label{sec:level2}Methodology}

\begin{figure*}
\includegraphics[width=\textwidth]{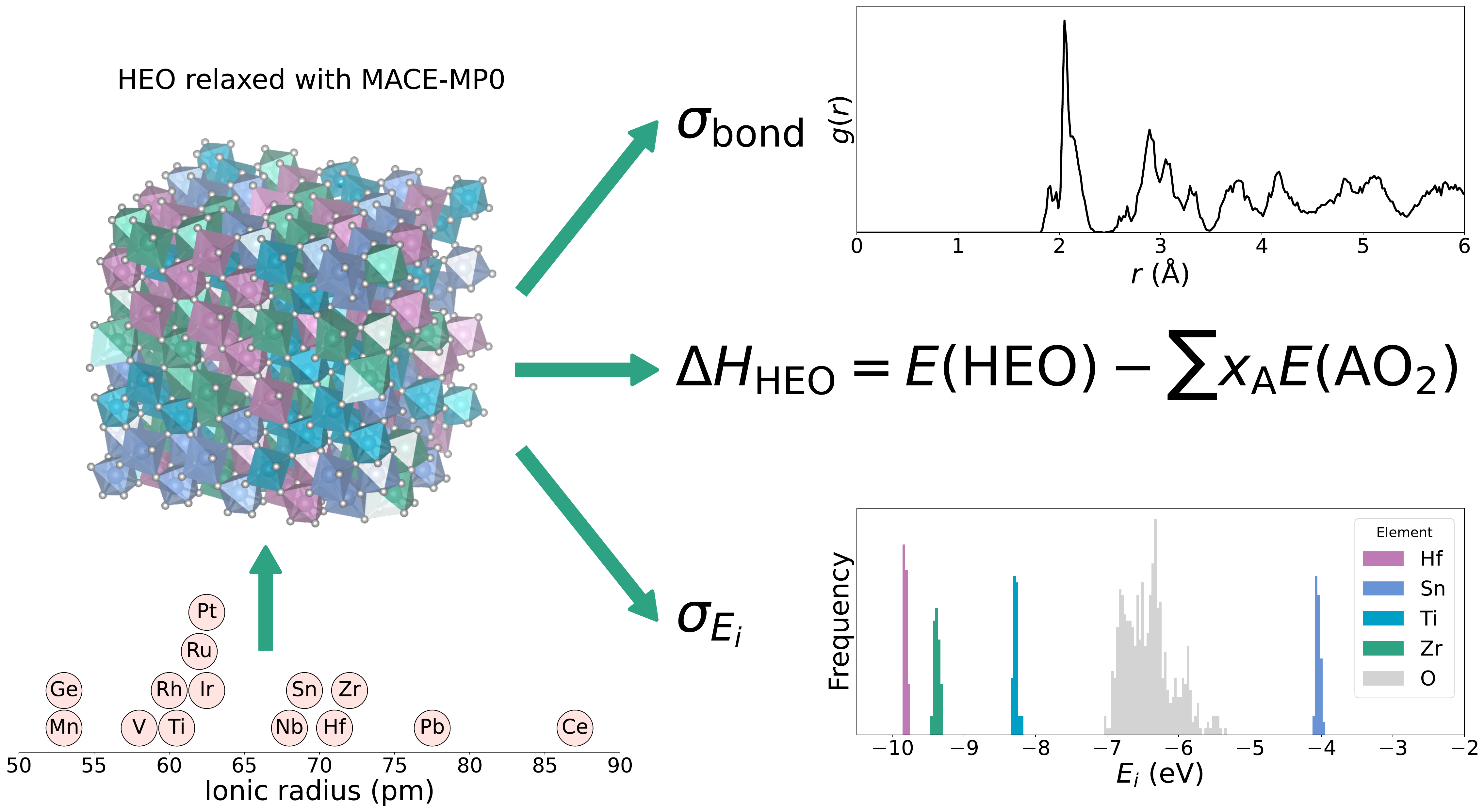}
\caption{\label{fig1:method_outline} An outline of the methodology for calculating synthesizability of tetravalent HEOs. First, a reasonably sized set of candidate structures must be chosen, in this case 14 tetravalent elements and three common crystal structures. Next, an HEO supercell with randomly distributed cations is generated and a structural relaxation is performed using the MACE machine learning interatomic potential. Finally, descriptors for entropy ($\sigma_{E_{i}}$), mixing enthalpy ($\Delta H_{\text{HEO}}$), and bond-lengths ($\sigma_{\text{bond}}$) are extracted. These descriptors are used to estimate the synthesizability of the HEOs.}
\end{figure*}

With the objective of screening thousands of candidates of 4- and 5-component tetravalent high entropy oxides, we have adapted and developed a descriptor-based methodology similar to previous HEO discovery searches \cite{Pitike2020, dey_mixed_2024, Divilov2024}. Comparable methodologies have previously been applied to tetravalent $A$O$_2$ compounds using DFT \cite{aamlid_phase_2023}, and for $A$O compounds using MLIPs \cite{sivak_discovering_2025}. The 14 elements Ti, V, Mn, Ge, Zr, Nb, Ru, Rh, Sn, Ce, Hf, Ir, Pt, and Pb are chosen due to their existence as tetravalent cations in $A$O$_2$ crystal structures, although this is not necessarily the most stable oxidation state for these elements depending on temperature and oxygen partial pressure. The four crystal structures \mbox{$\alpha$-PbO$_2$}, baddeleyite, rutile, and fluorite were selected to encompass the most probable tetravalent crystal structures.


\subsection{Calculating enthalpy of mixing using MACE}

Previous studies calculate descriptors in order to evaluate synthesizability. In all studies we have found \cite{Pitike2020, dey_mixed_2024, Divilov2024, sivak_discovering_2025}, the HEO's enthalpy of mixing is calculated or approximated and used as a descriptor. The prominence of this descriptor relies on the assumption that a low enthalpy of mixing can be overcome by the entropy gain from the configurational entropy of the target HEO at elevated temperatures, over a mixture of alternative competing phases that can be formed by the constituent elements. 

Here, we use the MACE-MP-0 foundation model \cite{batatia2023foundation} to relax and calculate the energies of high entropy oxide supercells, $E(\text{HEO})$, and the most stable binary oxide system for each cation at 0 K, $E(A\text{O}_{2})$ (described in full in section \ref{sec:oxidation}). All energies are normalized to one formula unit. The enthalpy of mixing of the HEO, $\Delta H_{\text{HEO}}$, is then calculated in a similar way to Sivak \emph{et al.} \cite{sivak_discovering_2025},
\begin{equation} \label{eq:mix_enthalpy}
    \Delta H_{\text{HEO}} = E(\text{HEO}) - \sum x_{A} E(A\text{O}_{2}), 
\end{equation}
where $x_A$ is the fraction of binary oxide $A\text{O}_{2}$. When the mixing enthalpy, $\Delta H_{\text{HEO}}$, is calculated using MACE or DFT it is denoted $\Delta H_{\text{MACE}}$ and $\Delta H_{\text{DFT}}$, respectively.

The HEO supercells are constructed by generating a supercell of $\sim$1000 atoms in a unit cell (see Fig. \ref{fig1:method_outline}) and randomly populating cation sites with the elements in the composition of interest in the correct ratios using the CLEASE code~\cite{chang_clease_2019}. This crystal is then relaxed using the MACE potential and the Atomic Structure Environment's (ASE) ExpCellFilter with the BFGS optimizer \cite{hjorth_larsen_atomic_2017}, which allows relaxation of the cell parameters and internal atomic positions.

A theoretical temperature of formation of the high entropy oxide from it's constituent binary oxides can be calculated from when the Gibbs free energy is equal to zero,
\begin{equation} \label{eq:free_energy}
    G = H - TS = 0,
\end{equation}
and thus the temperature of formation can be estimated as 
\begin{equation} \label{eq:mix_temp}
    T = \frac{\Delta H_{\text{HEO}}}{\Delta S_{\text{mix}}},
\end{equation}
where $\Delta S_{\text{mix}} = k_{\text{B}} \ln(N_{\text{cation}})$ is the ideal mixing entropy for an equimolar composition of $N_{\text{cation}}$ cation species. This serves as a lower bound for the temperature of formation, due to the simplistic assumption that the binary oxides react to form the HEO which is stabilized by the ideal configurational entropy, and does not take into account other competing phases or kinetic concerns. 

\subsection{Bond-length descriptor}
The bond-length descriptor proposed by Sivak \emph{et al.} \cite{sivak_discovering_2025} is defined as the standard deviation of the relaxed first nearest neighbour cation-oxygen bonds,
\begin{equation} \label{eq:bond_descriptor_Sivak}
    \tilde{\sigma}_{\text{bond}} = \sqrt{\frac{\sum_{i}|a_{i}-\overline{a}|^2}{N}},
\end{equation}
where $\overline{a}$ is the average nearest neighbour distance and $a_{i}$ is the distance of each cation-oxygen pair after relaxation.

This descriptor works well for characterizing rocksalts, where the cation-oxygen first nearest neighbors are all equidistant, and all cation sites are equivalent. However, this is not the case in many crystal structures, including those studied in this paper, where first shell oxygen-cation bond distances can be anisotropic, or when unit cells have multiple symmetry inequivalent cation sites. To enable the descriptor to be used for any crystal structure, we introduce a modified method for calculating the bond-length descriptor using the radial distribution function (RDF). After relaxation of the cell, the crystal's initial $\delta$-function peaks of the cation-oxygen RDF, $g_{\text{A-O}}(r)$, instead form a broad distribution reflecting disorder (see Fig.\ref{fig1:method_outline}). The bond-length descriptor then becomes the standard deviation of the first shell bond distances with respect to the average distance,
\begin{equation} \label{eq:bond_descriptor}
    \sigma_{\text{bond}} = \sqrt{\frac{\int_0^{r_{\text{cut}}}(r-\bar{r})^2n(r)dr}{\int_0^{r_{\text{cut}}}n(r)dr}},
\end{equation}
where $\bar{r}$ is the average A-O distance, $r_{\text{cut}}$ is the cutoff radius defined as the distance of the first minimum after the first RDF maximum (typically at approximately 2.5 \AA) and $n(r)$ is the radial atomic number density. This bond descriptor can be applied to HEOs of any crystal structure with the caveats that inherently anisotropic structures have larger baseline $\sigma_{\text{bond}}$ values, and that the RDF would have to be evaluated separately for crystal structures with multiple sublattices.

\subsection{Entropy descriptor}
We take advantage of the specific property of MLIPs such as MACE, where the total energy of a system is defined as the sum of known individual atom energies, $E_{\text{total}} = \sum_{i} E_{i}$, to build a novel entropy descriptor. These individual atom energies are a mapping from each atom's local environment and layers of its neighbours, and are thus influenced by more complex interaction energies and not simply the distances to its nearest neighbours. This entropy descriptor represents a physically motivated approach to approximate the entropy gain from a representation of the thermodynamic density of states as proposed by Divilov \emph{et al.}~\cite{Divilov2024}.

The standard deviation of each cation species ($A$) individual atom energies distribution is
\begin{equation} \label{eq:entropy_descriptor_A}
    \sigma^{A}_{E_{i}} = \sqrt{\frac{\sum_{i}|E^{A}_{i}-\overline{E^{A}_{i}}|^2}{N_{A}}}
\end{equation}
where $N_{A}$ is the number of cation $A$ in the system. The average of all single-cation $\sigma^{A}_{E_{i}}$ is then taken to calculate the entropy descriptor

\begin{equation} \label{eq:entropy_descriptor}
    \sigma_{E_{i}} = \frac{\sum_{A}\sigma^{A}_{E_{i}}}{N_{\text{cation}}} .
\end{equation}

The standard deviation of oxygen energies is not included as their distribution is not comparable between different chemical compositions. The distribution of oxygen atom energies is dominated by the spread of cation-oxygen pairs nearest-neighbour average energies (see Fig. \ref{fig1:method_outline}).

\subsection{Oxidation states}\label{sec:oxidation}

Although all the elements selected have stable tetravalent states, some are commonly found in other oxidation states and this must be considered when calculating the mixing enthalpy (see Eq.~\ref{eq:mix_enthalpy}). In order to calculate the lowest energy of the tetravalent binary oxide compounds, $E(A\text{O}_{2})$ according to the MACE potential, it is necessary to consider the binary oxides with different cation-oxygen ratios. This is achieved by calculating the relative energy of the tetravalent oxide with respect to other oxides with different valences and the energy of an oxygen atom (half the energy of an $\text{O}_2$ molecule), 
\begin{equation} \label{eq:binary_oxide}
    E(A\text{O}_{2}) = E(A_{x}\text{O}_{y}) + (y-2x)\frac{1}{2}E(\text{O}_{2})
\end{equation}
where $E(\text{O}_{2})$ is the energy of an isolated O$_2$ molecule calculated using MACE-MP-0 \cite{batatia2023foundation}, and $E(A_{x}\text{O}_{y})$ is the energy of a binary compound with formula $A_{x}\text{O}_{y}$. To find the lowest $E(A\text{O}_{2})$ for each element, all binary oxide compounds for that cation in the Materials Project database \cite{Jain2013, Horton2025} are calculated. 

\section{Results and Discussion}

\subsection{Comparison of MACE with DFT methods}\label{sec:MACEvsDFT}

\begin{figure*}[!tp]
\includegraphics[width=\textwidth,keepaspectratio]{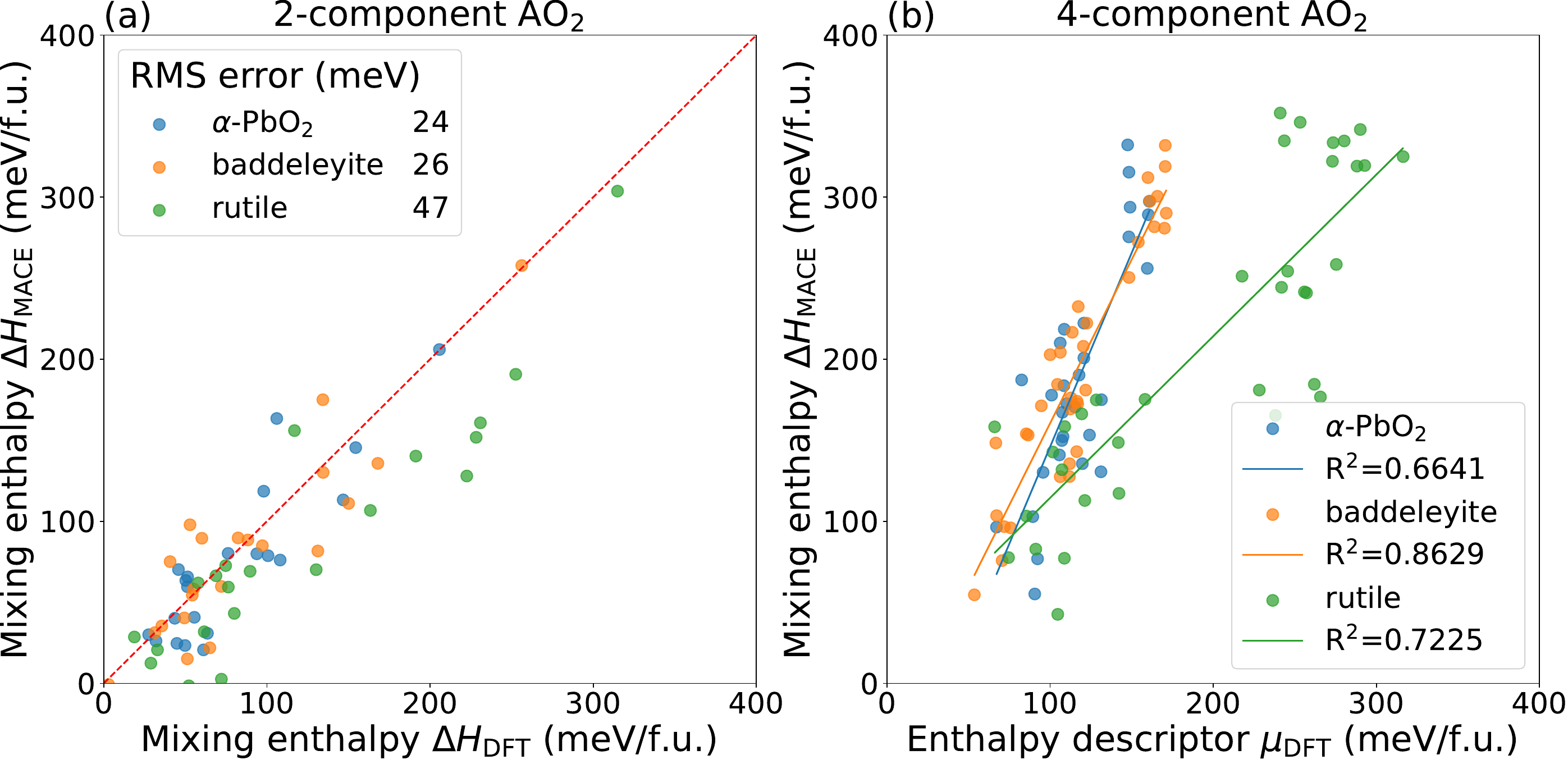}
\caption{\label{fig2:DFT_MACE_compare} (a) Comparison of the mixing enthalpies of 2-component $A$O$_2$ SQS cells calculated using MACE versus DFT, the dashed red line shows $\Delta H_{\text{MACE}}=\Delta H_{\text{DFT}}$. (b) Comparison of the mixing enthalpies of 4-component $A$O$_2$ calculated using MACE versus the simplified enthalpy descriptor $\mu_{\text{DFT}}$ suggested in Ref.~\cite{aamlid_phase_2023}.}
\end{figure*}

The use of MACE allows us to directly calculate the energies of thousands of $\sim$10$^3$ atom HEO systems at significantly reduced computational cost compared to DFT. Relaxing a single system of that size with DFT using a standard inexpensive functional would typically require approximately 1000 CPUs for 24 hours. MACE is capable of fully relaxing the same structure in approximately 1 minute on an NVIDIA Tesla V100. MACE boasts DFT-level accuracy \cite{batatia2023foundation} but it is not clear whether that accuracy translates to enthalpy of formation calculations for HEOs. HEOs have much more complex chemical environments than the crystals in the training set, but we will show that MACE performs well.

As can be seen in Fig.~\ref{fig2:DFT_MACE_compare}(a), there is a good agreement between MACE and DFT calculations of the 2-component compounds, ($A_{0.5}B_{0.5}$)O$_2$. In this case, MACE has been used to calculate the energies of all relaxed special quasirandom structure (SQS) cells of all pairwise combinations of the 7 elements Ti, Ge, Zr, Sn, Ce, Hf, Pb previously calculated using DFT~\cite{aamlid_phase_2023}, with no further relaxation of the SQS cells or binary oxides. Despite the DFT parameters not being exactly the same as those used to produce the training data which is collected from the Materials Project \cite{Jain2013, Horton2025}, the root mean square (RMS) error for the mixing enthalpies for different structures is 24-47 meV per formula unit. This is comparable to the mean absolute error (MAE) of 18.7 meV per atom for the total energies of rocksalt HEOs calculated previously using CHGNet and DFT \cite{sivak_discovering_2025}, or for the complete training set for MACE-MP-0 which has a MAE of 20 meV per atom, noting the normalization per atom rather than per formula unit \cite{batatia2023foundation}. Note that the data in Fig.~\ref{fig2:DFT_MACE_compare} is fitted to mixing enthalpies and not total energies, and is as such a measure of relative, rather than absolute, energies.

In order to search for tetravalent 4- and 5-component HEOs, some of the present authors previously calculated an enthalpy descriptor using DFT~\cite{aamlid_phase_2023}. This descriptor, $\mu_{\text{DFT}}$, is defined as the average of all pairwise, or 2-component, compound combinations' formation enthalpies, rather than a true mixing enthalpy. This was a necessary approximation as performing DFT calculations for large SQS cells of all possible 4- and 5-component HEOs was prohibitively expensive. However, MLIPs now allow us to directly calculate the mixing enthalpies of these 4- and 5-component structures by calculating the energies of large 500-1000 atom cells. To ensure the stability of the MACE potential and to ensure that the disorder inherent to HEOs is well sampled, we performed relaxations on 10 randomly generated structures of each compound and crystal structure and then took the average value of the formation enthalpy and entropy descriptor for the $\binom{7}{4} = 35$  4-component compounds. Given the typical RMS error over the 10 samples of 1-3 meV/f.u., only single structures were used later for the wider structure search.

As can be seen in Fig.~\ref{fig2:DFT_MACE_compare}(b), there is no direct agreement between the MACE calculated mixing enthalpy from Eq.~\ref{eq:mix_enthalpy}, $\Delta H_{\text{MACE}}$, and the DFT enthalpy descriptor, $\mu_{\text{DFT}}$, of the 4-component systems, though there is a positive correlation, with the line of best fit for rutile having a gradient of 1.00, $\alpha$-PbO$_2$ 2.39 and baddeleyite 2.02. It should be emphasized that $\mu_{\text{DFT}}$ is a descriptor where the average mixing enthalpy of 10 (for the 4-component HEOs) or 15 (for the 5-component HEOs) pairwise interaction energies is not directly equivalent to a calculation of the full 4-component system. A major effect might be that of the volume relaxation, which is done independently for each cation pair DFT calculations, as opposed to globally for the whole system in the MACE calculations. This could explain why $\mu_{\text{DFT}}$ underestimates the enthalpy of mixing compared to the MACE calculations. $\mu_{\text{DFT}}$ does, however, correlate with $\Delta H_{\text{MACE}}$, suggesting that $\mu_{\text{DFT}}$ can still be suitable for ordering of compounds by enthalpy of mixing, although the absolute value cannot be directly contrasted with the mixing entropy as in Eq.~\ref{eq:mix_temp}. The MLIPs include more complex interactions than just 2-body terms and have an equivalent computational cost, allowing for a better description of the complete system.

\subsection{Effectiveness of descriptors for structure prediction}\label{sec:effectiveness}

\begin{figure*}
\includegraphics[width=\textwidth,keepaspectratio]{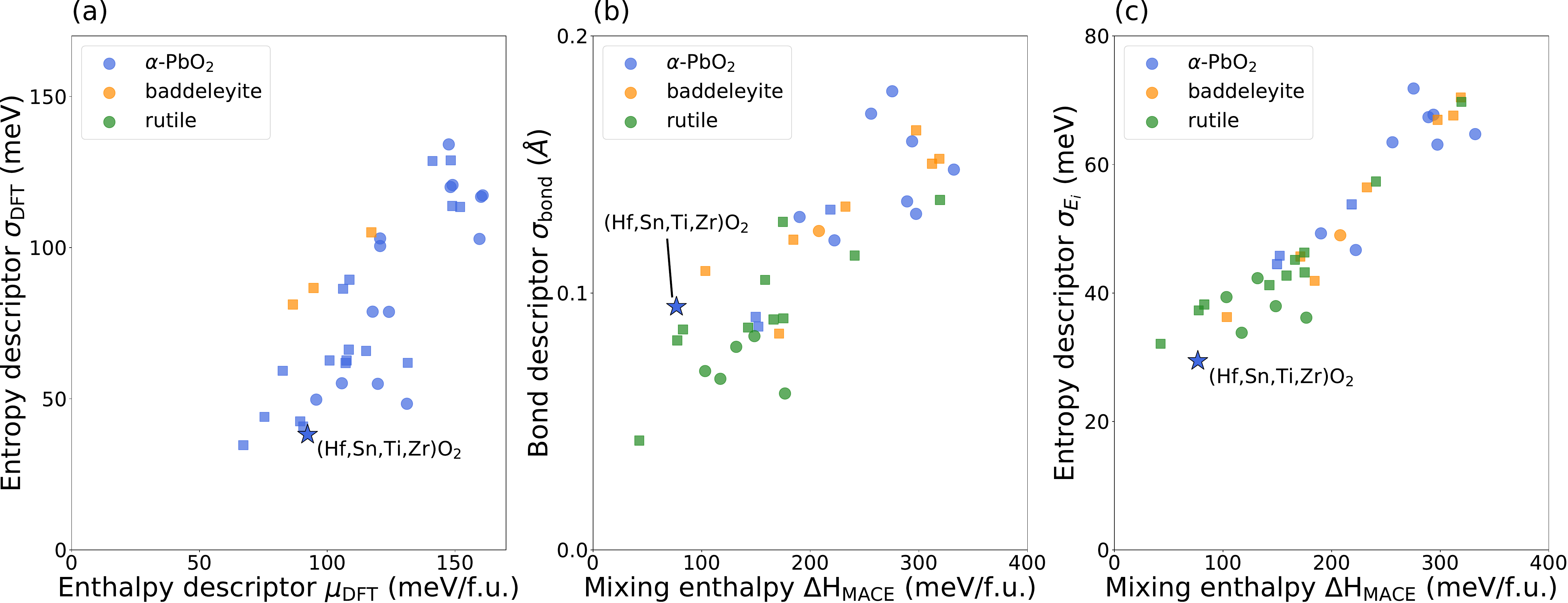}
\caption{\label{fig3:Hvdescriptor} Scatter diagrams of the 35 4-component compounds 
where only the crystal structure with the lowest entropy descriptor ($\sigma_{\text{DFT}}$ or $\sigma_{E_{i}}$) for each compound is shown. The star indicates the only successfully synthesized compound, (Hf,Sn,Ti,Zr)O$_2$, which all approaches correctly find to prefer the $\alpha$-PbO$_2$ structure, and all Pb-containing compounds are shown by a square.
(a) Entropy descriptor, $\sigma_{\text{DFT}}$, plotted against the enthalpy descriptor, $\mu_{\text{DFT}}$, both calculated previously using DFT \cite{aamlid_phase_2023}. The successfully synthesized compound is identified disregarding the Pb-containing compounds.
(b) The MACE calculated enthalpy of mixing, $\Delta H_{\text{MACE}}$, plotted against the bond length descriptor, $\tilde{\sigma}_{\text{bond}}$. By this metric, the successfully synthesized compound is not identified as the most promising.
(c) The MACE calculated enthalpy of mixing, $\Delta H_{\text{MACE}}$, plotted against the MACE calculated $E_{i}$ entropy descriptor, $\sigma_{E_{i}}$. The successfully synthesized compound is clearly identified as the most promising candidate.
}
\end{figure*}

To computationally predict the synthesizability of undiscovered HEOs, one would ideally calculate the temperature-dependent free energy of the $(\!N_{\text{cation}}\!+\!1\!)$-dimensional phase diagram of all the component elements and their potential structures, following the CALPHAD method \cite{Spencer2008}. Even utilizing MLIPs, this is computationally expensive to compute for a single HEO, and unfeasible for hundreds of candidate compounds.
To limit the computational cost, descriptor-based methods as detailed in Section \ref{sec:level2} are used. The mixing enthalpy (Eq.~\ref{eq:mix_enthalpy}) is the most commonly used descriptor for HEO prediction. This descriptor is typically calculated as the relative enthalpy of formation of the disordered HEO with respect to the mixture of component binary oxides ($A$O$_2$) at 0 K. 

There are some limitations of $\Delta H_{\text{HEO}}$ as a descriptor. Firstly, the enthalpy of the binary oxides is temperature dependent, due to various factors including the change in oxygen chemical potential at higher temperatures and the change in volume and bond distances of the constituent oxides. This temperature dependence is most notable for Pb which begins to reduce from PbO$_2$ to PbO at approximately 300$^{\circ}$C under ambient conditions, thus requiring a correction to be applied which raises $\Delta H_{\text{HEO}}$. That correction is not applied here to be consistent with the existing data presented, but it should be noted that when the correction is applied, Pb-containing compounds tend to become considerably less favorable \cite{aamlid_phase_2023}.

A second limitation of $\Delta H_{\text{HEO}}$ as a descriptor is that this enthalpy of mixing only considers the binary constituents, while alternative competing phases (ordered or disordered mixtures of the ternary, quaternary etc. oxides) that may be more energetically favorable are not calculated. This partly explains why formation enthalpy cutoffs for synthesizability are often stated as being approximately 100 meV/f.u. \cite{aamlid_phase_2023,sivak_discovering_2025} despite furnaces easily achieving temperatures of 1500$^{\circ}$C, equivalent to a mixing enthalpy of 210 or 244 meV/f.u. for a 4- or 5-component HEO, respectively (from Eqn. \ref{eq:mix_temp}). For screening purposes, the cutoff enthalpy for synthesizability is lower than the ideal theoretical maximum, most likely due to the complications of polymorphs, ternary ordered phases, or partially disordered phases, that can form at high temperatures.

Fig.~\ref{fig3:Hvdescriptor} shows the performance of three different pairs of descriptors used to describe the 35 4-component compounds. Only the data point of the crystal structure with the lowest entropy descriptor is shown for each chemical composition. The star signifies the known synthesizable compound, (Hf,Sn,Ti,Zr)O$_2$, while the squares signify Pb-containing compounds to highlight issues with Pb reduction. The leftmost panel, Fig.~\ref{fig3:Hvdescriptor}(a), summarizes the results from \cite{aamlid_phase_2023}. The enthalpy descriptor, $\mu_{\text{DFT}}$, is defined as the average of all the constituents pairwise mixing enthalpies, as discussed in Sec. \ref{sec:MACEvsDFT}. The entropy descriptor, $\sigma_{\text{DFT}}$, is the variance of all the constituents pairwise mixing enthalpies. If an oxidation penalty on Pb is applied the successfully synthesized compound is more clearly distinguished, as it has the lowest value of both enthalpy and entropy descriptor.

In Fig.~\ref{fig3:Hvdescriptor}(b) the bond length descriptor, $\tilde{\sigma}_{\text{bond}}$, is plotted against the MACE calculated enthalpy of mixing, $\Delta H_{\text{MACE}}$. Here the crystal phase shown is the same as in Fig.~\ref{fig3:Hvdescriptor}(c) for easier comparison. If the lowest bond descriptor is used to determine the most favorable crystal phase instead, it predicts all phases, except one, to be rutile, including (Hf,Sn,Ti,Zr)O$_2$. This highlights the inability of the bond length descriptor to fairly evaluate inherently anisotropic coordination environments, such as the one found in the $\alpha$-PbO$_2$ structure.

In this work, we introduce a new descriptor, $\sigma_{E_i}$ (see Eq.~\ref{eq:entropy_descriptor}), that uses the standard deviation of the individual atom energies of the constituent cations as a measure of entropy. In Fig.~\ref{fig3:Hvdescriptor}(c), this new descriptor is plotted against the MACE calculated enthalpy of mixing, $\Delta H_{\text{MACE}}$. The successfully synthesized compound is clearly identified in the correct crystal structure, without additional considerations about Pb. The use of both these descriptors predicts more potential compounds in the rutile structure over the $\alpha$-PbO$_2$ structure, but $\sigma_{E_i}$ correctly predicts (Hf,Sn,Ti,Zr)O$_2$ to form in the $\alpha$-PbO$_2$ phase. As none of the other compounds have been successfully synthesized in the lab it could well be that the rutile structure is the most favourable for many of these compounds.

\begin{figure}
\includegraphics[width=\linewidth,keepaspectratio]{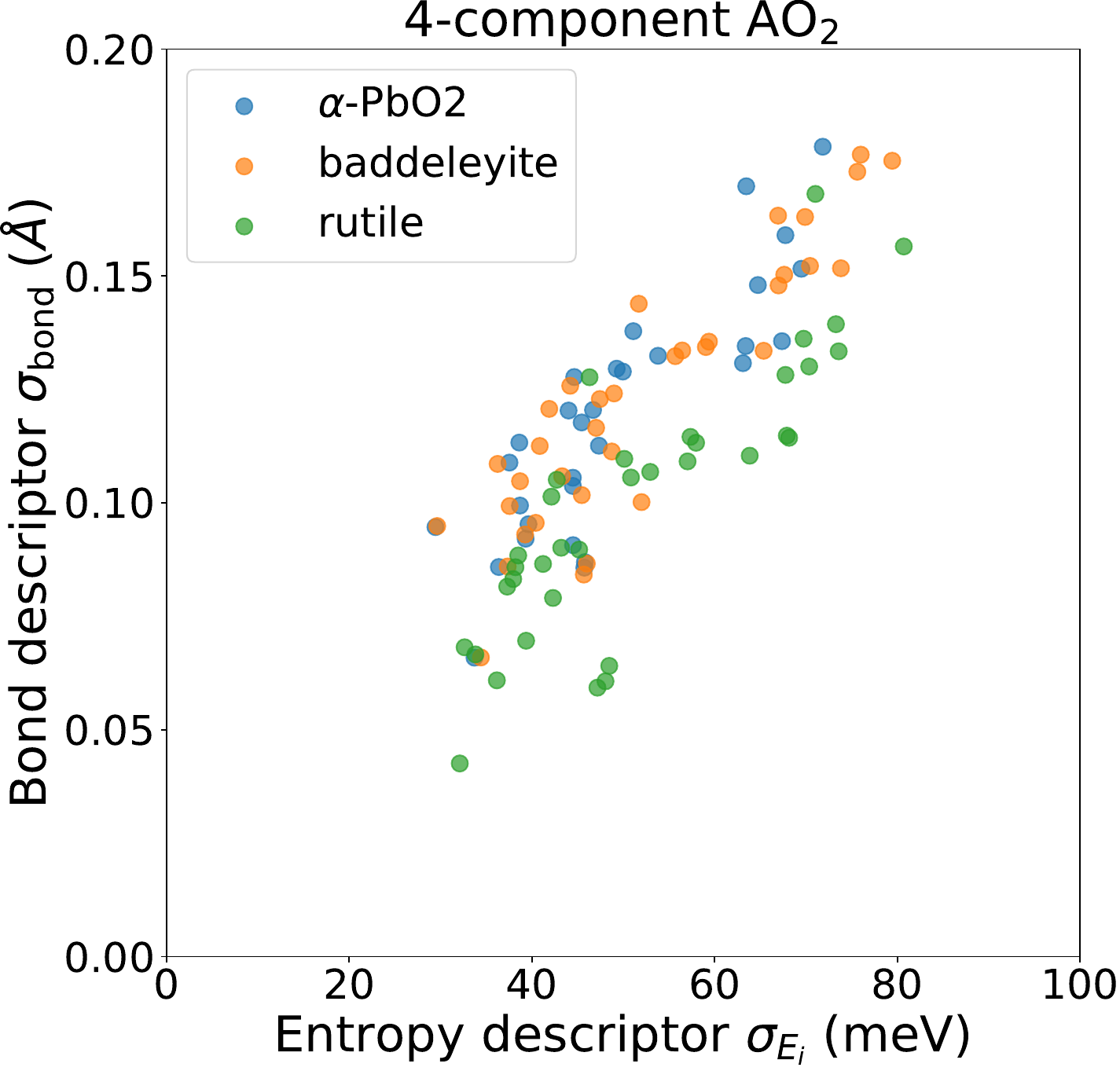}
\caption{\label{fig4:sigma_compare} A comparison between the two descriptors, the bond-length descriptor $\tilde{\sigma}_{\text{bond}}$ and the individual atom energy based entropy descriptor, $\sigma_{E_{i}}$, from the MACE calculations for each 4-component compound and crystal structures.}
\end{figure}

As can be seen in Fig.~\ref{fig4:sigma_compare}, our newly introduced descriptor does correlate with the bond-length descriptor previously used in rock-salts to predict synthesizability \cite{sivak_discovering_2025}. This suggests that bond-descriptors measure similar characteristics of the relaxed structures to the entropy descriptor we have proposed. However, while the entropy descriptor does broadly correlate with structural disorder, if we compare Fig.~\ref{fig3:Hvdescriptor}(b) and (c), it becomes clear that the entropy descriptor is better able to distinguish the one successfully synthesized tetravalent HEO compound, (Hf,Sn,Ti,Zr)O$_2$, from the other calculated compounds. The use of the bond-descriptor would suggest that there are other synthesizable compounds with both lower $\Delta H$ and $\sigma_{\text{bond}}$. For the purposes of screening, whilst use of $\sigma_{\text{bond}}$ would allow us to narrow down the number of compositions to attempt in the lab, the entropy descriptor provides finer resolution, distinguishing compounds with similar bond disorder which still have unfavorable local environments. Furthermore, the synthesized compound falls within the low $\Delta H$ and low $\sigma_{E_i}$ region, indicating its superior utility as a predictor of synthesizability.

\subsection{High-throughput synthesizability search of 4- and 5- component compositions}

We applied our methodology to an extensive search over 4- and 5-component HEO compositions drawn from a 14-element set (Ti, V, Mn, Ge, Zr, Nb, Ru, Rh, Sn, Ce, Hf, Ir, Pt, and Pb), requiring the calculation of 3,003 unique chemical compositions. Each composition was evaluated in three candidate structures: \mbox{$\alpha$-PbO$_2$}, baddeleyite, and rutile. Initial calculations of the compounds in the fluorite phase had significantly higher enthalpies of formation and so were discounted. In total this required the full relaxation of approximately 10,000 supercells of $\sim$10$^3$ atoms.

\begin{figure}[!tp]
\includegraphics[height=0.8\textheight
]{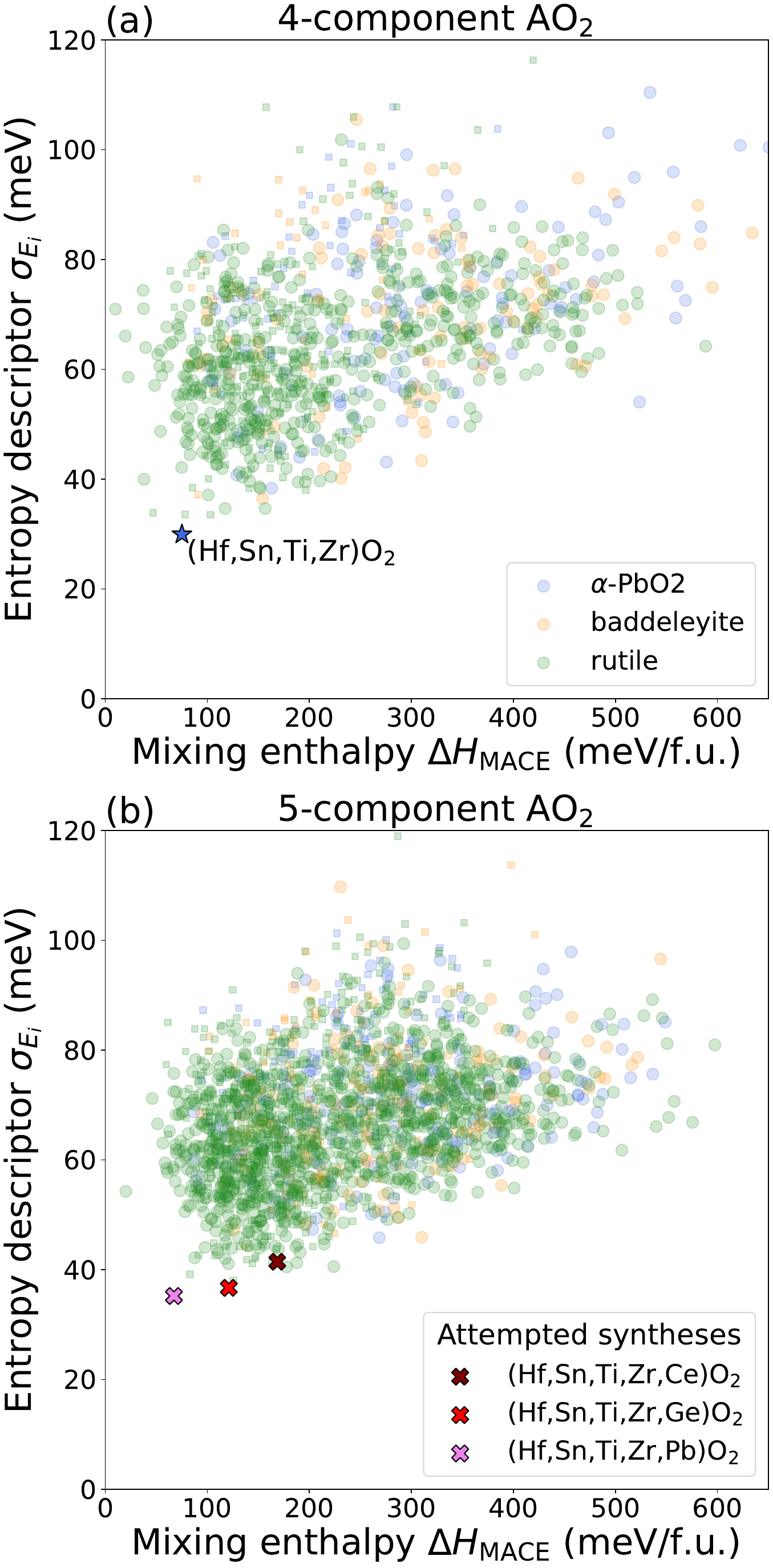}
\caption{\label{fig5:4and5component} Scatter plots of all (a) 4- and (b) 5-component $A$O$_2$ HEO compounds showing the entropy descriptor $\sigma_{E_{i}}$ versus the enthalpy of mixing $\Delta H_{\text{MACE}}$ for the relaxed crystal phases with the lowest synthesizability descriptor, $\rho$. All compounds containing Pb are shown as smaller squares to highlight the lack of applied Pb-penalty. The successfully synthesized $\text{(Hf,Sn,Ti,Zr)O}_{2}$ compound is shown as a star (blue denoting it is in the $\alpha$-PbO$_2$ phase), and the unsuccessful synthesis attempts~\cite{aamlid_phase_2023} are shown as crosses. 
}
\end{figure}

With the main objective of this high-throughput method being to discover new HEOs, a measure of synthesizability is required. For a compound to be synthesizable, both the enthalpy descriptor and the entropy descriptor have to be sufficiently low. A large enthalpy descriptor would lead to the formation of competing phases, while a large entropy descriptor indicates that certain interactions are more or less favorable than others, leading to formation of ordered phases or clustering. An estimate of synthesizability based on both these descriptors is essentially a closeness to zero criterion, where the importance given to the enthalpy and entropy descriptors can be weighted. 

A couple of synthesizability descriptors have been suggested in the literature. A mixed enthalpy-entropy descriptor has been proposed and demonstrated on high entropy metal carbides~\cite{dey_mixed_2024}. Their enthalpy and entropy descriptors are normalized based on the search space, which works well for rock-salt structured carbides due to the prevalence of high entropy compositions in this space, but is not as applicable to a search space where discoveries are expected to be rarer. A disordered enthalpy–entropy descriptor, where one descriptor is divided by the other has been applied to carbides, carbonitrides, and diborides \cite{Divilov2024}. However, the boundary between synthesizable and not is dependent on the system under investigation.

We suggest a different estimate of synthesizability 
\begin{equation} \label{eq:MEED}
    \rho = \frac{1}{\sqrt{2}}\sqrt{\left(\frac{ \sigma_{E_{i}}}{w_{\sigma}} \right)^2 +\left(\frac{ \Delta H_{\text{HEO}}}{w_{H}} \right)^2}
\end{equation}
where $w_{\sigma}$ and $w_{H}$ are scaling factors for the entropy and enthalpy descriptor terms respectively. Scaling factors are required in order to define the relevant "distance" metric for each descriptor. There are many potential ways one could set the scaling factors, either from the distribution of the calculated data, or knowledge from lab-based synthesis. In this case, as there is only one known example of a successfully synthesized compound, (Hf,Sn,Ti,Zr)O$_2$, we set its synthesizability to 1.0 by setting $w_{\sigma}$ and $w_{H}$ equal to the compound's calculated entropy and enthalpy descriptor values. This would then suggest anything with a synthesizability lower than 1.0 has a high probability of being made in the lab, with probability decreasing as the synthesizability value increases.

Figure \ref{fig5:4and5component} displays a scatter plot of $\sigma_{E_i}$ vs. $\Delta H_{\text{MACE}}$ for the crystal phase with the lowest synthesizability descriptor, $\rho$, for all (a) 4- and (b) 5-component compounds. The rutile structure (green) dominates as the most likely crystal phase according to $\rho$, accounting for 70$\%$ and 75$\%$ of predicted 4- and 5-component compounds respectively. Notably, even in this dramatically expanded composition space, the known synthesizable 4-component system (Hf,Sn,Ti,Zr)O$_2$ is clearly distinguished in the region of low entropy and enthalpy descriptor, and is also predicted to be in the correct phase, $\alpha$-PbO$_2$. Only Pb-containing compounds (shown as squares) have values near it, as a physically motivated reduction penalty is not applied~\cite{aamlid_phase_2023}. This supports our hypothesis that synthesizability correlates with both low energetic disorder and thermodynamic stability. It also corroborates the experimental experience that there may only be a small fraction of HEO compositions, when compared to the multitude possible, that may be synthesizable or metastable at high temperatures. 

The only other compounds with a relatively low $\rho$ that do not contain Pb (the full list ordered by $\rho$ is available in the Supplementary Material~\cite{MySuppMat}) are predicted to be rutile, with the lowest being (Ir,Pt,Rh,Ru)O$_2$ and (Hf,Pt,Sn,Ti)O$_2$. The compound with the second lowest synthesizability descriptor predicted to have the $\alpha$-PbO$_2$ crystal structure is (Ge,Hf,Ti,Zr)O$_2$, similar to the only successfully synthesized compound with Ge swapped in for Sn. However, this compound appears at 50th position in the list ranked according to $\rho$ and has a relatively high mixing enthalpy of 129 meV/f.u. and entropy descriptor of 41 meV/f.u. This shows further evidence of the selectivity of the descriptor, where it correctly predicts the $\alpha$-PbO$_2$ structure of (Hf,Sn,Ti,Zr)O$_2$.

The high-throughput search is extended to 5-component systems in Fig.~\ref{fig5:4and5component}(b). The attempted syntheses from ref.~\cite{aamlid_phase_2023} are highlighted, each of which involve the same 4 elements as the known synthesizable compound with the addition of 1 element. These syntheses failed, in air, for three different reasons: in the case of (Hf,Sn,Ti,Zr,\textbf{Ce})O$_2$ the mixing enthalpy is too high, in the case of (Hf,Sn,Ti,Zr,\textbf{Pb})O$_2$ a reduction of Pb is observed, and in the case of (Hf,Sn,Ti,Zr,\textbf{Ge})O$_2$ an ordered zircon-structured phase is formed.

The broader trend across all compositions suggests that these combined descriptors effectively prioritize candidate systems. If we assume that $\rho$ is not an absolute cutoff, several new 5-component systems can be tentatively identified as targets for synthesis, potentially extending the known compositional range of tetravalent HEOs. However, several compounds that were previously found not to be synthesizable via solid state synthesis are in close proximity to those targets. Only four of the calculated 5-component HEOs have a lower $\rho$ than (Hf,Sn,Ti,Zr,\textbf{Ge})O$_2$. Two of them are also additions to the successfully synthesized 4-component HEO, namely (Hf,Sn,Ti,Zr,\textbf{Pt})O$_2$ and (Hf,Sn,Ti,Zr,\textbf{Mn})O$_2$. The other two are similar to one of the novel predicted 4-component compounds, (Ir,Mn,Pt,Rh,Ru)O$_2$ and (Hf,Sn,Ti,Pt,Rh)O$_2$. These compounds are all predicted to be rutile by the synthesizability descriptor.

Given the thousands of calculated structures, the fact that so few compounds are near the synthesizability region shows once more the nontrivial nature of the search for stable HEOs.

\subsection{Discussion}

The speed-up that comes from using the MACE foundation model to screen thousands of compositions and structures per day cannot be overstated. The alternative is to run far fewer, simplified, DFT calculations, or use brute force guided by chemical intuition in a laboratory setting to attempt to synthesize new materials from a much reduced sampling pool. However, there are some limitations and blindspots to the MLIP approach.

A fundamental limitation of MACE is its lack of explicit descriptors for long-range interactions, relying on the Atomic Cluster Expansion (ACE) \cite{drautz_atomic_2019} to describe local environments and inference from DFT training data for longer-range effects.
MACE-MP-0 is trained on the Materials Project Trajectory (MPtrj) Dataset \cite{Deng2023figshare}, which contain information on chemical compositions, structures, energies, magnetic moments, forces, and stresses. While this enables DFT-comparable accuracy, large regions of chemical and structural space remain sparsely represented, occasionally leading to non-converging calculations or unphysical atomic relaxations.

Fortunately, the field of universal MLIPs is progressing quickly. Additional training data from open sources such as the Alexandria dataset \cite{schmidt_machine-learning-assisted_2023} and Open Materials 2024 \cite{barroso-luque_open_2024} are now available. MACE-MPA-0, a new version, includes the Alexandria dataset and halves the MAE of MACE-MP-0. New machine learning models are easy to insert into the methodology pipeline, sometimes only requiring changing one line of code, and any improvements in potentials can be quickly taken advantage of.

There are other promising avenues for improving model accuracy without relying on new foundational models trained at great expense on large datasets. Fine-tuning MACE on targeted data, especially for chemical environments that lie on the fringes of the training distribution, has been shown to improve convergence and stability in previously problematic systems \cite{kaur_data-efficient_2025}. Building these small, targeted, datasets can be achieved by using methods like active learning, which use uncertainty quantification \cite{bilbrey_uncertainty_2025} to select structures in areas of the models maximum uncertainty and so limit the number of expensive DFT calculations necessary.

Polymorphism, where the same chemical composition can exhibit a variety of crystal structures, is a common behavior in oxides and poses a challenge for MLIPs. TiO$_2$ has 12 polymorphs, with enthalpy differences as small as a few meV/f.u., meaning DFT struggles to rank phases reliably dependent on the functionals and pseudopotentials used \cite{Curnan2015}. Similarly, in (Hf,Sn,Ti,Zr)O$_2$ the enthalpy difference between the $\alpha$-PbO$_2$ and rutile structures calculated with MACE is only $\sim 4$ meV/f.u.. Nevertheless, if lab synthesis yields an unexpected polymorph, it can still be viewed as a predictive success.

Oxides can exhibit multiple oxidation states depending on temperature and oxygen partial pressure. The selection of elements explored here contains some noble metals (Ru, Rh, Ir, Pt) which will reduce to metals at intermediate to high temperatures unless exposed to highly oxidizing conditions. Some of these elements are also present in compounds with low $\rho$ values. On the contrary, oxides of V and Nb will oxidize at intermediate to high temperatures unless they are sealed in a practically oxygen-free atmosphere. Mixing elements from these two groups may prove difficult, due to the incompatibility of the oxygen chemical potentials necessary to keep the oxidation state at 4+. A solution to this issue is proposed in using a oxygen chemical potential overlap descriptor~\cite{almishal2025thermodynamics}.

Unlike DFT, MACE has no explicit information about charge states and electronic bands; all energetics are inferred from the local structural and chemical environment. However, the charge states are to some extent inferred by the neural net from the local environments. An attempt to include Sc, which is nominally trivalent, in the calculations led to issues where the model predicted compounds with unreasonable oxidation states, even when correcting for oxidation potentials as outlined in Sec.~\ref{sec:oxidation}. The training data does not have sufficient information about nominally tetravalent Sc, and as such does not know that it is unfavorable. 

There are additional experimental limitations to synthesis that the calculations also do not take into account. For instance, if the melting point of some of the components is much lower than others, it might be impossible to find a suitable temperature for solid state synthesis where none of the components are melting, yet all have sufficient thermal energy to diffuse and react. Volatilization is also an issue with certain reactants. Such limitations can partially be overcome by ensuring that there is a feedback loop between computational predictions and experimental synthesis, so that the results can be filtered based on practical know-how of which combinations are reasonable. Depending on the specific practical problem, alternative synthesis methods might be able to circumvent the issues~\cite{chen2019mechanochemical,gonzalez2024impact}.

\section{Conclusions and outlook}
The potential chemical composition space for high entropy oxides is enormous, and yet discovery of synthesizable compounds are relatively rare and experimentally driven. In this work, we have demonstrated a computationally efficient and physically grounded method for computational prediction of synthesizability in high entropy oxides, by applying it to a model system comprised of tetravalent elements and crystal structures.

Leveraging the universal MACE machine learning interatomic potential, we evaluated the enthalpy of mixing, local bond disorder, and a novel energy-based entropy descriptor across thousands of candidate compositions and structures. Our comparison with DFT data and prior experimental results confirms the validity of MACE predictions for multi-component disordered systems. We showed that while the enthalpy descriptor provides a general indication of stability, our entropy descriptor captures additional effects which are equally critical to synthesizability.

This framework provides a blueprint for large-scale screening of disordered oxide systems and paves the way toward rational design of synthesizable high entropy materials. Extending to new search spaces with different elements, crystal structures, and even to off-equimolar compositions is straightforward. Computational work in this field has, up until now, been focused on rationalizing and testing existing families of compounds. We hope that in the near future we will find true computational predictions of new high entropy materials using this method.

\section{Data availability}

Data and code is available at \url{https://github.com/od-qmul/HEO\_search}.

\begin{acknowledgments}
This work was funded by the UKRI Guarantee Marie Skłodowska-Curie grant administered through the Engineering and Physical Sciences Research Council (EP/X034429/1).
We are grateful to our colleagues at the Quantum Matter Institute’s “Atomistic approach to emergent properties of disordered materials” Grand Challenge for insightful conversations on high entropy oxides. We would also like to thank Prof. Christoph Ortner and Cheuk Hin Ho
for their advice on MLIPs.
This research was undertaken thanks in part to funding from the Natural Sciences and Engineering Research Council of Canada (NSERC) and the Canada First Research Excellence Fund (CFREF), Quantum Materials and Future Technologies Program.
\end{acknowledgments}

\bibliography{aPbO2_heo}

\end{document}